\documentclass[12pt]{arXiv_class}
\usepackage{authblk}
\usepackage{url}
\usepackage{amsmath}
\newcommand\EatDot[1]{}

\usepackage[misc]{ifsym}
\usepackage{listings}
\usepackage{jlcode}

\newcommand\blfootnote[1]{%
  \begingroup
  \renewcommand\thefootnote{}\footnote{#1}%
  \addtocounter{footnote}{-1}%
  \endgroup
}

\title{\textbf{StateSpaceModels.jl: a Julia Package for Time-Series Analysis in a State-Space Framework}}

\date{}

\author{Raphael Saavedra, Guilherme Bodin, and Mario Souto}

\begin{document}

\maketitle

\begin{abstract}

\small{\texttt{StateSpaceModels.jl}} is an open-source Julia package for modeling, forecasting and simulating time series in a state-space framework. The package represents a straightforward tool that can be useful for a wide range of applications that deal with time series. In addition, it contains features that are not present in related commercial software, such as Monte Carlo simulation and the possibility of setting any user-defined linear model.


\end{abstract}
\blfootnote{
	
	Raphael Saavedra (\Letter~rsaavedra@ele.puc-rio.br), Guilherme Bodin, and Mario Souto are with the Department of Electrical Engineering and the Laboratory of Applied Mathematical Programming and Statistics (LAMPS), Pontifical Catholic University of Rio de Janeiro (PUC-Rio), Rio de Janeiro, Brazil.
	
	This work was supported in part by the Energisa Group through R\&D project ANEEL PD-00405-1701/2017.
}
\section{Motivation} \label{sec:introduction}

State-space modeling is a classical framework in control engineering that represents a system through the definition of input, state, and output variables \cite{zadeh2008linear}. Input variables are external entities that are inserted into the system and can serve as control inputs or noise. State variables represent unobserved components that evolve through time following a given state equation and also depending on the values of the input variables. Finally, output variables result from the realization of the state plus noise factors and represent the observable outcome of the system. In general, state-space models make use of the Kalman filter \cite{kalman1960new, welch1995introduction} to obtain predictive estimates for the state.

Due to its comprehensive form and wide range of potential applications, state-space models found a niche in time-series modeling, forecasting, and simulation, representing a flexible framework for time-series analysis with time-varying parameters \cite{durbin2012time}. The ability to conveniently define the evolution of a time series allows the characterization of stochastic components, such as trend and seasonality, that are non-trivial to model in other frameworks.

There are several packages focused on state-space models for time-series analysis in other languages, such as \texttt{KFAS} in R \cite{kfas} and \texttt{Statsmodels} in Python \cite{seabold2010statsmodels}, as well as commercial software such as \texttt{STAMP} \cite{koopman2000stamp}. Among related Julia packages, we highlight:
\begin{itemize}
    \item ~\texttt{Kalman.jl} \cite{kalmanjl}, which implements Kalman filtering and smoothing.
    \item ~\texttt{StateSpace.jl} \cite{statespacejl}, a control-oriented package with several versions of the Kalman filter.
    \item ~\texttt{ControlSystems.jl} \cite{controlsystemsjl}, a package aimed at control applications with numerous features, including state-space modeling.
\end{itemize}

The contributions brought by these packages notwithstanding, there are currently no state-space model packages aimed at time-series analysis in Julia. Furthermore, some have not been updated for a long time and are not functional in Julia 1.0. Thus, the main objectives of \texttt{StateSpaceModels.jl} \cite{statespacemodels} are:

\begin{enumerate}
    \item to fill this gap by implementing a general and intuitive framework in Julia for modeling, estimating, forecasting and simulating time series with state-space models;
    \item to provide an open-source package that is capable of performing the same functions as related commercial software such as \texttt{STAMP} and more;
    \item to develop a package for time-series analysis with state-space models that is fully implemented in Julia, contrary to related packages in other languages that depend on C or Fortran \mbox{routines} \cite{kfas}.
\end{enumerate}

The remainder of this paper is organized as follows. In Section \ref{sec:statespace}, the Gaussian state-space framework is introduced. In Section \ref{sec:modeling}, the procedures for model specification are presented. Section \ref{sec:estimation} explains the filtering, estimation and smoothing processes. In Section \ref{sec:forecasting}, the forecasting and simulation procedures are presented. Section \ref{sec:applications} contains examples of applications that illustrate the use of the package. Finally, conclusions and future work are provided in Section \ref{sec:conclusion}.

\section{Gaussian state-space framework} \label{sec:statespace}

For the sake of consistency and readability, we utilize in the package the same notation as \cite{durbin2012time}. Suppose we have a series of observations $y_{1}, \dots, y_{n}$. In a state-space framework, it is assumed that the observations depend on a set of unobserved components denominated states, or simply the state, and denoted by $\alpha_{1}, \dots, \alpha_{n}$. The state components often have physical interpretations, such as trend and seasonality in a time-series context or position and speed in a control setting.

The main idea behind state-space modeling is to define the evolution of the state and its relation with the observed variables. This is done through the following two equations, denoted the observation equation and the state equation, respectively:
\begin{align}
    y_{t} &= Z_{t} \alpha_{t} + \varepsilon_{t}, \hspace{0.045\columnwidth} \varepsilon_{t} \sim N(0, H_{t}), \label{eq:obs_equation} \\
    \alpha_{t+1} &= T_{t} \alpha_{t} + R_{t} \eta_{t}, \quad \eta_{t} \sim N(0, Q_{t}), \label{eq:state_equation}
\end{align}
\noindent where $y_{t}$ is a $p \times 1$ vector of observations and $\alpha_{t}$ is an unobserved $m \times 1$ vector representing the state at instant $t$. Note that the behavior of the system over time is determined by $\alpha_{t}$ as defined in \mbox{Eq. \eqref{eq:state_equation}}, but $\alpha_{t}$ cannot be directly observed, contrary to $y_{t}$.

Matrices $Z_{t}$, $T_{t}$, and $R_{t}$ are the ones that define how the observations relate to the state and how the state evolves over time, and are generally assumed to be known. The error terms $\varepsilon_{t}$ and $\eta_{t}$ are supposed to be serially independent and independent of each other. Finally, $H_{t}$ and $Q_{t}$ are the covariance matrices of the error terms. In this package, it is assumed that $T$, $R$, $H$, and $Q$ are not time-varying, as this is the case for the vast majority of practical applications. Conversely, $Z_{t}$ is allowed to vary over time.

Next, for illustration purposes, we present an example model and show how it can be inserted into the state-space framework. Consider the following model, called the linear trend model:
\begin{align}
    y_{t} &= \mu_{t} + \varepsilon_{t}, \hspace{0.067\columnwidth} \varepsilon_{t} \sim N(0, \sigma^{2}_{\varepsilon}), \label{eq:lineartrend1} \\
    \mu_{t+1} &= \mu_{t} + \nu_{t} + \xi_{t}, \quad \xi_{t} \sim N(0, \sigma^{2}_{\xi}), \label{eq:lineartrend2} \\
    \nu_{t+1} &= \nu_{t} + \zeta_{t}, \hspace{0.07\columnwidth} \zeta_{t} \sim N(0, \sigma^{2}_{\zeta}), \label{eq:lineartrend3}
\end{align}
\noindent where $\mu_{t}$ represents the trend component and $\nu_{t}$ the slope component. Note that the we can rewrite this model in the framework of Eq. \eqref{eq:obs_equation}--\eqref{eq:state_equation} in the following manner:
\begin{align}
    y_{t} &= \begin{bmatrix} 1 & 0 \end{bmatrix} \alpha_{t} + \varepsilon_{t}, \hspace{0.105\columnwidth} \varepsilon_{t} \sim N(0, \sigma^{2}_{\varepsilon}), \\
    \alpha_{t+1} &= \begin{bmatrix} 1 & 1 \\ 0 & 1 \end{bmatrix} \alpha_{t} + \begin{bmatrix} 1 & 0 \\ 0 & 1 \end{bmatrix}\!\eta_{t}, \quad \eta_{t} \sim N \bigg(\!\begin{bmatrix} 0 \\ 0 \end{bmatrix}\!, \begin{bmatrix} \sigma^{2}_{\xi} & 0 \\ 0 & \sigma^{2}_{\zeta} \end{bmatrix}\!\bigg).
\end{align}
Similarly, more complex and sophisticated models can also be modeled in the state-space framework.

\section{Model specification} \label{sec:modeling}

Model specification is the step wherein the user defines the state-space model they want to consider. This is done through the creation of a \texttt{StateSpaceModel} structure that contains the observations $y_{t}$ as well as matrices $Z_{t}$, $T$, and $R$.

\subsection{Predefined models} \label{subsec:predefined}

A set of predefined classical models is available, namely the local level model, the linear trend model, the basic structural model, and the structural model with exogenous variables. These models can be conveniently defined by calling a function and providing the observations $y_{t}$. Note that all models automatically extend to the multivariate case.

\subsubsection{Local level model}

The local level model consists of a stochastic level component that is defined by a random walk:
\begin{align}
    y_{t} &= \mu_{t} + \varepsilon_{t}, \quad \varepsilon_{t} \sim N(0, \sigma^{2}_\varepsilon), \label{eq:locallevel1} \\
    \mu_{t+1} &= \mu_{t} + \xi_{t}, \quad \xi_{t} \sim N(0, \sigma^{2}_\xi). \label{eq:locallevel2}
\end{align}
A local level \texttt{StateSpaceModel} can be created using:
\begin{lstlisting}[language = Julia]
model = local_level(y)
\end{lstlisting}
\noindent where \texttt{y} is an observation vector in the univariate case or a matrix in column-wise fashion in the multivariate case, i.e., each column representing a variable and each line representing a time period.

\subsubsection{Linear trend model}

The linear trend model consists of a local level model with the addition of a stochastic slope term, as seen in Eq. \eqref{eq:lineartrend1}--\eqref{eq:lineartrend3}. A linear trend \texttt{StateSpaceModel} can be created using:
\begin{lstlisting}[language = Julia]
model = linear_trend(y)
\end{lstlisting}

\subsubsection{Structural model}

The basic structural model consists of three stochastic components: trend, slope, and seasonality, defined in the following manner:
\begin{align}
    y_{t} &= \mu_{t} + \gamma_{t} + \varepsilon_{t}, \hspace{0.095\columnwidth} \varepsilon_{t} \sim N(0, \sigma^{2}_\varepsilon), \label{eq:structural1} \\
    \mu_{t+1} &= \mu_{t} + \nu_{t} + \xi_{t}, \hspace{0.095\columnwidth} \xi_{t} \sim N(0, \sigma^{2}_\xi), \label{eq:structural2} \\
    \nu_{t+1} &= \nu_{t} + \zeta_{t}, \hspace{0.145\columnwidth} \zeta_{t} \sim N(0, \sigma^{2}_\zeta), \label{eq:structural3} \\
    \gamma_{t+1} &= \sum_{j=1}^{s-1} \gamma_{t+1-j} + \omega_{t}, \hspace{0.06\columnwidth} \omega_{t} \sim N(0, \sigma^{2}_\omega). \label{eq:structural4}
\end{align}
In this model, there is a stochastic trend that contains a slope component, similarly to the linear trend model. Furthermore, a stochastic seasonal component is modeled so that its sum over the seasonality period $s$ is equal to zero except for a noise factor. A basic structural \texttt{StateSpaceModel} can be created using:
\begin{lstlisting}[language = Julia]
model = structural(y, s)
\end{lstlisting}

Additionally, a structural model can contain exogenous variables, also known as explanatory variables. Exogenous variables represent external factors that are correlated to the phenomenon in question and can, thus, be used to improve the estimation and forecasting by increasing the available information. This model is identical to \eqref{eq:structural1}--\eqref{eq:structural4} except for the observation equation, which has an additional exogenous factor:
\begin{align}
    y_{t} &= \mu_{t} + \gamma_{t} + \theta^{\top} X_{t} + \varepsilon_{t}, \quad \varepsilon_{t} \sim N(0, \sigma^{2}_\varepsilon) \label{eq:structural_exogenous}
\end{align}
A structural model with exogenous variables can be created with:
\begin{lstlisting}[language = Julia]
model = structural(y, s; X = X)
\end{lstlisting}
\noindent where \texttt{X} is a matrix containing the exogenous variables observations in column-wise fashion.

\subsection{User-defined models}

Apart from the well-known predefined models presented in Section \ref{subsec:predefined}, the package allows the input of any user-defined linear model through the definition of matrices $Z_{t}$, $T$, and $R$ through the following constructor:
\begin{lstlisting}[language = Julia]
model = StateSpaceModel(y, Z, T, R)
\end{lstlisting}

In the simpler case where $Z_{t}$ is constant, the argument \texttt{Z} can be input as an \texttt{Array\{Float64, 2\}}. Conversely, if $Z_{t}$ is time-varying, it must be input as an \texttt{Array\{Float64, 3\}} where the third dimension represents the time periods.

\section{Filtering, estimation, and smoothing} \label{sec:estimation}

Next, the model needs to be estimated. In this step, values for the covariance matrices $H$ and $Q$, initially assumed to be unknown, are obtained via maximum likelihood estimation \cite{casella2002statistical}. In order to do that, the Kalman filter needs to be utilized. Additionally, a smoother is applied in order to obtain the so-called smoothed estimates for the state.

After specifying the model, the estimation step can be called with the function \texttt{statespace}:
\begin{lstlisting}[language = Julia]
ss = statespace(model; verbose = 1)
\end{lstlisting}

In this step, the Kalman filtering, the maximum likelihood estimation of the fixed parameters, and the smoothing are conducted. The optional argument \texttt{verbose}, which defaults to 1, specifies the verbosity: 0 for no output, 1 for some progress information, and 2 for progress information as well as the optimization log.

An example of the estimation log with default verbosity can be seen below.

\begin{minipage}[adjusting]{0.52\columnwidth}
\begin{lstlisting}[language = Julia]
julia> ss = statespace(model)
==============================================================
                  StateSpaceModels.jl v0.2.0
 (c) Raphael Saavedra, Guilherme Bodin, and Mario Souto, 2019
--------------------------------------------------------------
            Starting state-space model estimation.
    Initiating maximum likelihood estimation with 3 seeds.
--------------------------------------------------------------
             Seed 0 is aimed at degenerate cases.
--------------------------------------------------------------
||    seed    |     log-likelihood      |      time (s)     ||
||       0    |        -16217.4939      |          0.21     ||
||       1    |         -1350.4763      |          2.65     ||
||       2    |         -1350.4763      |          4.08     ||
||       3    |         -1350.4763      |          5.69     ||
--------------------------------------------------------------
           Maximum likelihood estimation complete.
                  Log-likelihood: -1350.4763
             End of state-space model estimation.
==============================================================
\end{lstlisting}
\end{minipage}

\subsection{Filtering}

The filtering step derives the predictive and filtered states as well as their covariance matrices at every time period via the Kalman filter. The predictive state and its variance are given by
\begin{align}
    a_{t+1} &= \mathbb{E}[\alpha_{t+1} | Y_{t}], \\
    P_{t+1} &= \mathbb{V}[\alpha_{t+1} | Y_{t}],
\end{align}
while the filtered state and its variance are given by
\begin{align}
    a_{t|t} &= \mathbb{E}[\alpha_{t} | Y_{t}], \\
    P_{t|t} &= \mathbb{V}[\alpha_{t} | Y_{t}],
\end{align}
where $Y_{t}$ denotes the set of time periods $y_{1}, \dots, y_{t}$.

Furthermore, the filter also computes the innovations, or prediction errors, as well as their covariance matrix at every time period:
\begin{align}
    v_{t} &= y_{t} - Z a_{t}, \\
    F_{t} &= \mathbb{V}[v_{t}].
\end{align}

The package contains an implementation of the standard Kalman filter and the square-root Kalman filter \cite{durbin2012time}. The latter is a variant which utilizes Cholesky decomposition to ensure that the computed covariance matrices are positive semidefinite, thus avoiding numerical errors.

Additionally, in order to provide a flexible framework, we allow the utilization of any user-implemented variant of the Kalman filter through an abstract type called \texttt{AbstractFilter}:
\begin{lstlisting}[language = Julia]
struct MyKalmanFilter <: AbstractFilter
\end{lstlisting}

The choice of the filter type must be done in the estimation step as follows:
\begin{lstlisting}[language = Julia]
ss = statespace(model; filter_type = SquareRootFilter)
\end{lstlisting}

If no \texttt{filter\_type} is provided, the standard type \texttt{KalmanFilter} is utilized as a default option. Further information on Kalman filtering can be found in Chapter 4.3 of \cite{durbin2012time}.

\subsection{Estimation}

The Kalman filter is responsible for deriving estimates for the predictive state and for its covariance matrix at each time period, but it is dependent on the values of the noise covariances $H$ and $Q$. Thus, these constant parameters need to be estimated via maximum likelihood \cite{casella2002statistical}. The log-likelihood function is given by
\begin{align}
    \ell(Y_{n}) = -\frac{np}{2} \log 2 \pi - \frac{1}{2} \sum_{t=1}^{n} (\log |F_{t}| + v_{t}^{\top} F_{t}^{-1} v_{t}),
\end{align}
which depends on the innovations and their covariance matrix at each time period. Therefore, at each optimization iteration, the Kalman filter needs to be executed.

The implemented optimization method \texttt{RandomSeedsLBFGS} generates random initial values for the parameters and then uses the L-BFGS algorithm \cite{liu1989limited} to maximize the log-likelihood function via the unconstrained optimization package \texttt{Optim.jl} \cite{mogensen2018optim}.

Additionally, similar to the filtering step, users are able to define any desired optimization method through abstract type \texttt{AbstractOptimizationMethod}:
\begin{lstlisting}[language = Julia]
struct MyOptimizationMethod <: AbstractOptimizationMethod
\end{lstlisting}

As with the filter type, the choice of optimization method must be done in the estimation step:
\begin{lstlisting}[language = Julia]
ss = statespace(model; optimization_method = RandomSeedsLBFGS())
\end{lstlisting}

More advanced parameters, such as optimizer tolerances and the number of seeds, can be set within the \texttt{RandomSeedsLBFGS} constructor.

\subsection{Smoothing}

The smoothing step is conducted after estimation is complete. It consists in obtaining the so-called smoothed state and its covariance matrix at each time period, i.e.,
\begin{align}
    \hat{\alpha}_{t} &= \mathbb{E}[\alpha_{t} | Y_{n}], \\
    V_{t} &= \mathbb{V}[\alpha_{t} | Y_{n}].
\end{align}
Similar to the filtering step, the smoothing is conducted through an iterative process. The procedure is done backwards, starting from the final time period $n$. The smoothed state estimates are useful for analyzing the behavior of the state components. Further information on state smoothing can be found on Chapter 4.4 of \cite{durbin2012time}.

\subsection{Missing observations}

One interesting property of the Kalman filter and smoother is their ability to treat missing observations. Suppose we have a time series spanning time periods $t = 1, \dots, n$, but for some periods $\tau, \dots, \tau^{*}$ we have no observations available. This represents an obstacle for estimating a model in several frameworks. Nonetheless, the Kalman filter and smoother allow the derivation of minimum variance linear unbiased estimates for the missing observations so that a completion of the time series is possible. Conveniently, the only necessary modification in the filtering and smoothing equations is considering $Z_{t} = 0$ for the missing period.

In the package, this is implemented so that any \texttt{NaN} values in the observations are considered missing observations and automatically treated in the filtering and smoothing steps. In this way, the estimation is successfully conducted even when the time series of interest is incomplete.

\begin{figure}[h]
	\centering
	\includegraphics[width=0.7\columnwidth]{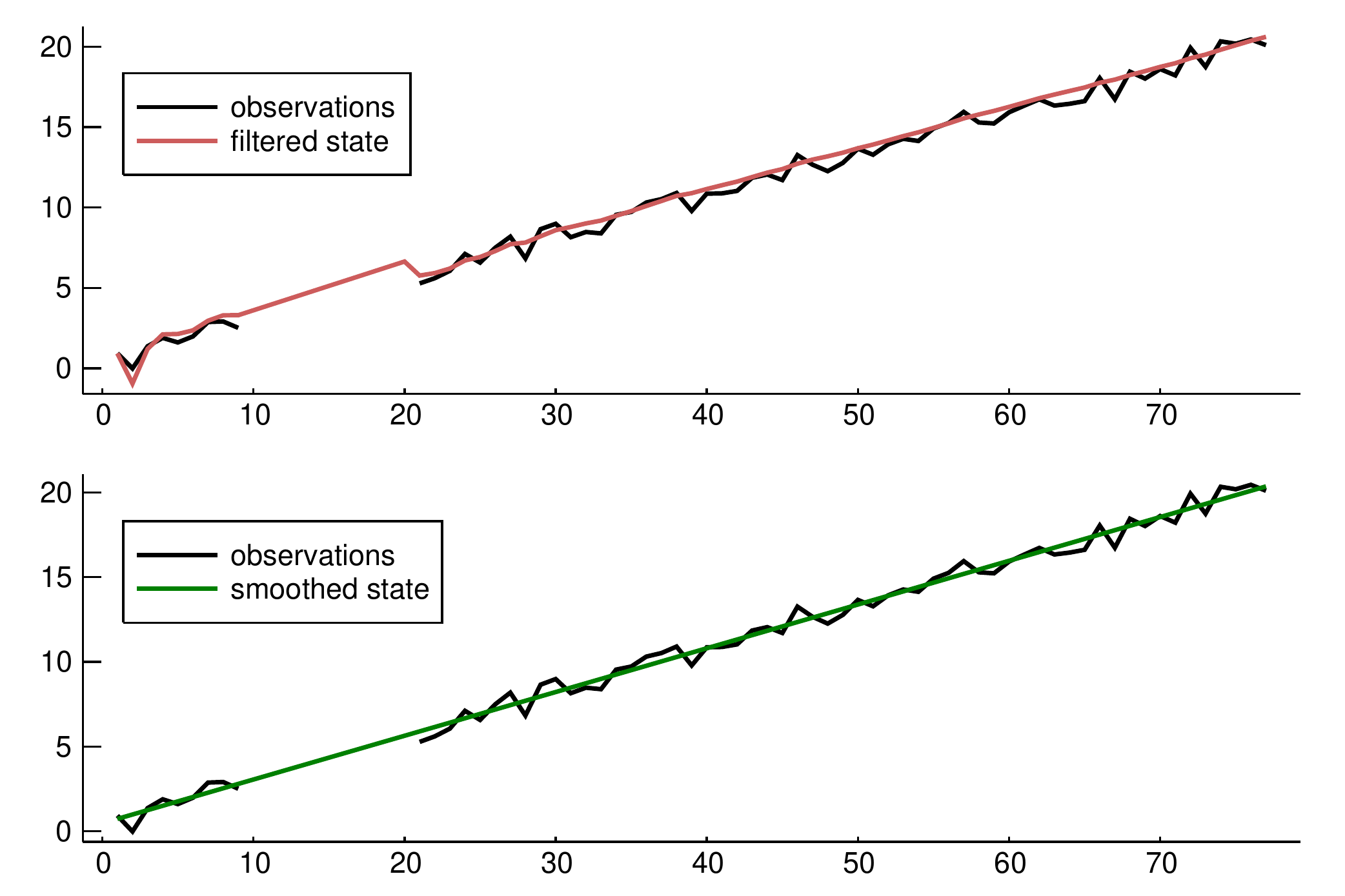}
	\caption{Automatic completion of missing observations in the filtering and smoothing.}
	\label{fig:missing_observations}
\end{figure}

Consider the following simulated time series composed by a growing trend plus a Gaussian noise. Additionally, we remove observations 10 through 20 to treat them as missing:
\begin{lstlisting}[language = Julia]
# Create growing trend series with Gaussian noise
y = collect(1:0.25:20) + 0.5*randn(77)
# Remove observations 10 through 20
y[10:20] .= NaN
# Specify the state-space model and estimate it
model = linear_trend(y)
ss = statespace(model)
\end{lstlisting}
The results of the estimation step are presented in Fig. \ref{fig:missing_observations}. The predictive, filtered, and smoothed state estimates can be accessed in \texttt{ss.filter.a}, \texttt{ss.filter.att}, and \texttt{ss.smoother.alpha}, \mbox{respectively}. It is clear in Fig. \ref{fig:missing_observations} that the filter and smoother are able to effectively capture the growing linear trend of the series even in the period where observations are missing.

\subsection{Estimation outputs}

At the end of the filtering, estimation, and smoothing routine executed by the \texttt{statespace} function, a \texttt{StateSpace} structure is returned. It contains six fields:

\begin{itemize}
    \item ~\texttt{model}: the previously defined \texttt{StateSpaceModel}.
    \item ~\texttt{filter}: a \texttt{FilterOutput} structure representing the result of the Kalman filter. It contains the predictive and filtered states as well as the innovations and their covariance matrices at each time period. Furthermore, there is a flag indicating if steady state was attained and the time period when it was attained.
    \item ~\texttt{smoother}: a \texttt{SmoothedState} structure representing the result of the smoothing procedure. It contains the smoothed state as well as its covariance matrix at each time period.
    \item ~\texttt{covariance}: a \texttt{StateSpaceCovariance} structure which contains the covariance matrices of the observation and state noises.
    \item ~\texttt{filter\_type}: the defined filter type.
    \item ~\texttt{optimization\_method}: the defined optimization method.
\end{itemize}

\section{Forecasting and Simulation} \label{sec:forecasting}

Finally, after the model has been estimated, forecasting and Monte Carlo simulation can be conducted. In the case of forecasting, estimates for the future values of the time series are computed along with their probability distributions. Alternatively, it is possible to simulate an arbitrary number of future scenarios of the series, which is useful for a wide range of applications such as stochastic optimization.

The minimum mean square error forecasts for $y_{t}$ can be obtained by treating future values of $y_{t}$ as missing values. Forecasting is conducted with the function \texttt{forecast}, which receives a \texttt{StateSpace} structure coming from the estimation step and the number of time periods ahead to be forecast, and outputs the minimum square error forecasts and the predictive distributions of $y_{t}$ at each time period:
\begin{lstlisting}[language = Julia]
pred, dist = forecast(ss, N)
\end{lstlisting}
Alternatively, another powerful tool is the simulation of future scenarios, since these are used as input by several applications, such as certain stochastic optimization models. Performing Monte Carlo simulation in a time-series state-space framework involves sampling several scenarios for the observation and state errors utilizing the estimated variances $H$ and $Q$ at each time period, and then computing the state-space recursions for each scenario.

Monte Carlo simulation can be conducted with the use of the function \texttt{simulate}, which receives a \texttt{StateSpace} structure coming from the estimation step, the number of time periods ahead $N$, and the number of scenarios $S$ to be simulated, and outputs an $N \times S$ matrix of scenarios for $y_{t}$:
\begin{lstlisting}[language = Julia]
scenarios = simulate(ss, N, S)
\end{lstlisting}

\section{Applications} \label{sec:applications}

In this section, we present several practical applications which can be addressed via state-space modeling. We utilize \texttt{StateSpaceModels.jl} to tackle the presented problems and provide the results. Furthermore, for the sake of reproducibility, the code of some of the examples is in the folder \texttt{examples} in the \texttt{StateSpaceModels.jl} repository.

\subsection{Airline passengers}

\begin{figure}[h]
	\centering
	\includegraphics[width=0.7\columnwidth]{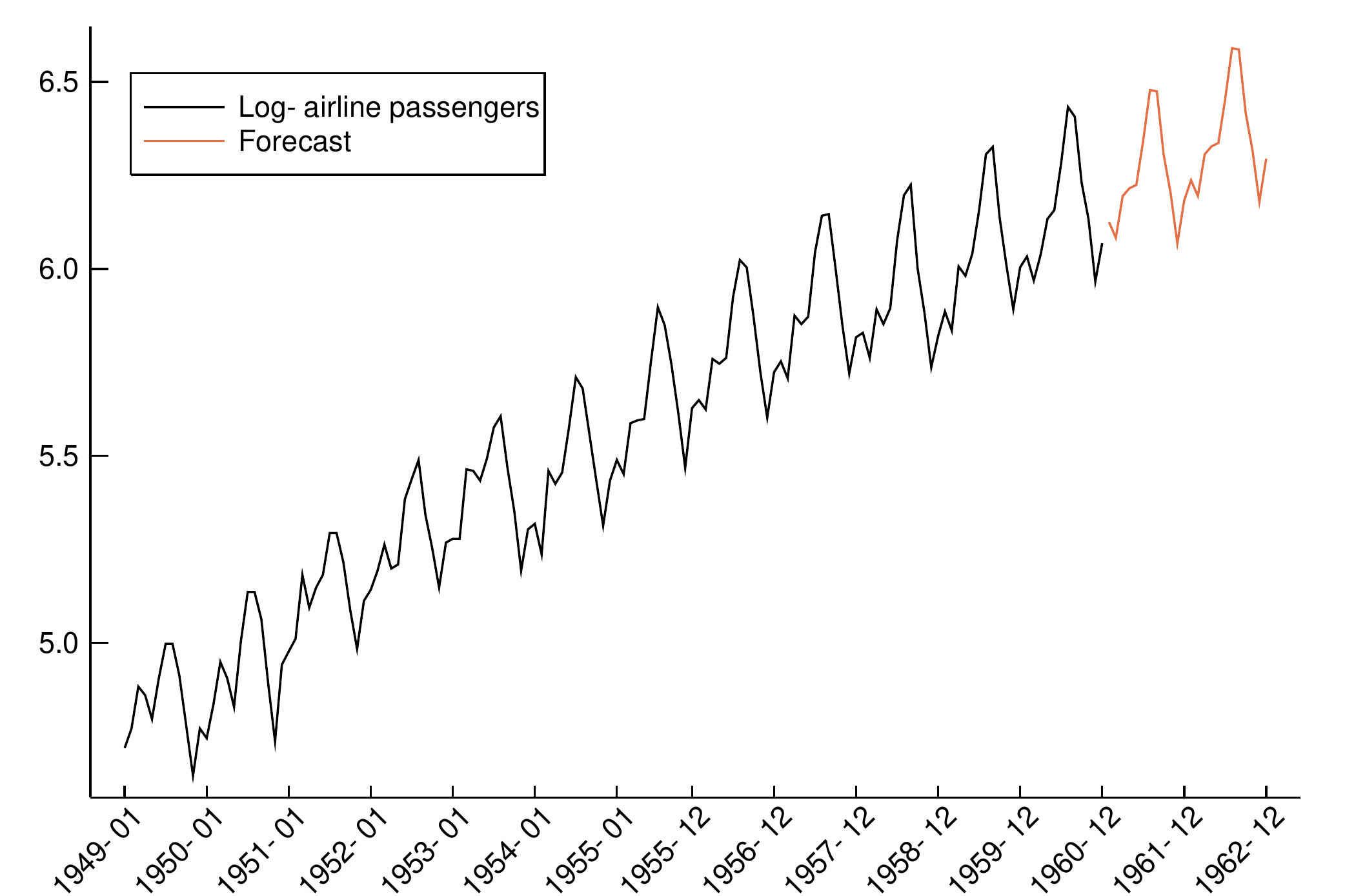}
	\caption{Log-airline passengers historical data and forecasting.}
	\label{fig:log_airline_forecast}
\end{figure}

As a first example, let us use the classical monthly airline passengers time series. In order to avoid multiplicative effects, we use the well-known approach of taking the log of the series. Fig. \ref{fig:log_airline_forecast} shows the log-airline passengers time series. We can estimate a structural \texttt{StateSpaceModel} as follows.
\begin{lstlisting}[language = Julia]
# Load the AirPassengers dataset
AP = CSV.read("AirPassengers.csv")
# Take the log of the series
logAP = log.(Vector{Float64}(AP[:Passengers]))
# Specify the state-space model
model = structural(logAP, 12)
# Estimate the state-space model
ss = statespace(model)
\end{lstlisting}

\begin{figure}[h]
	\centering
	\includegraphics[width=0.7\columnwidth]{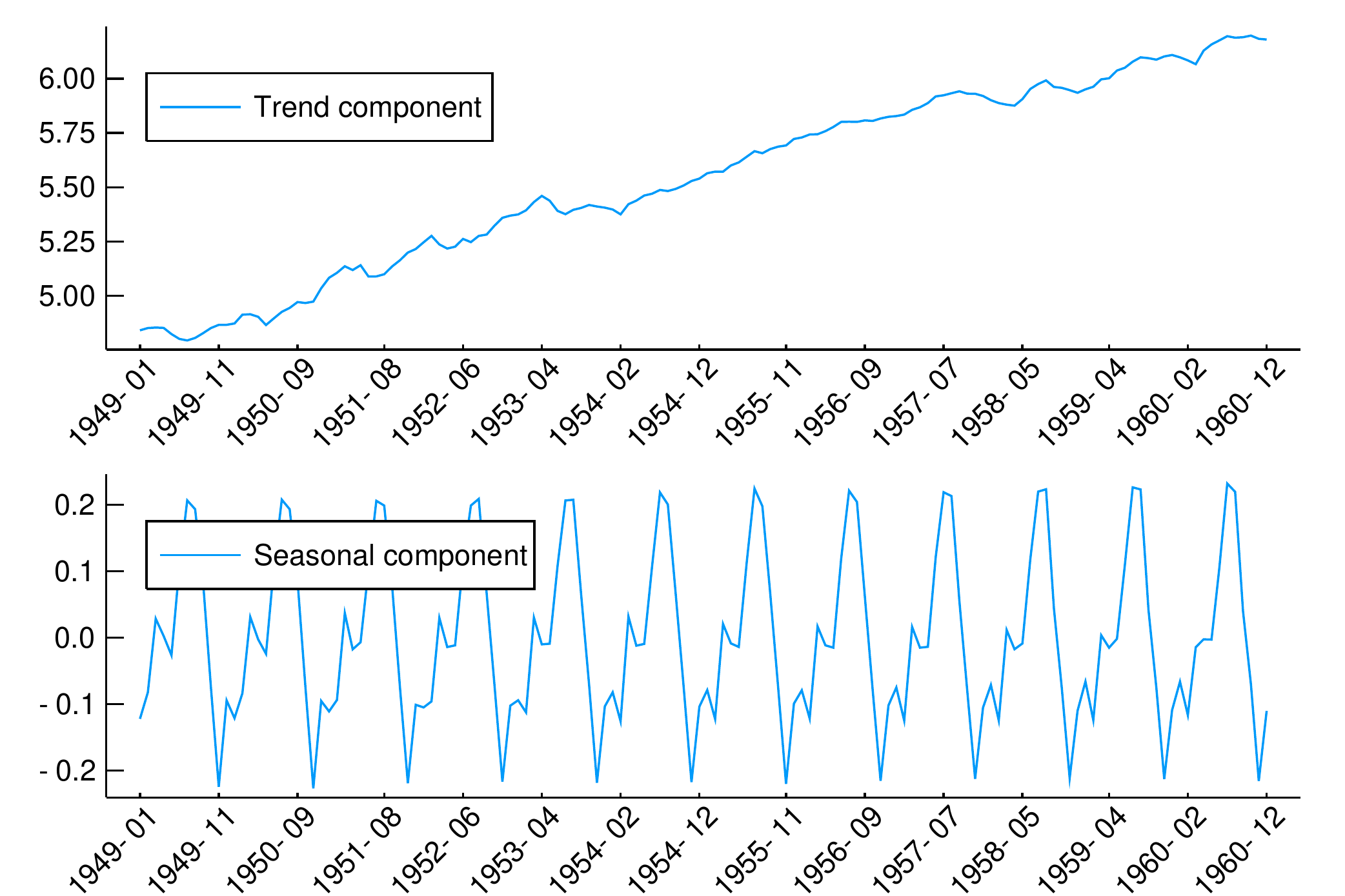}
	\caption{Smoothed trend and seasonal components of the log-airline passengers time series.}
	\label{fig:log_airline_components}
\end{figure}

By estimating a structural model, we can analyzed the individual components of the series, such as trend and seasonality. These components are presented in Fig. \ref{fig:log_airline_components} and represent dimensions 1 and 3 of the smoothed state, respectively.

We can also forecast the following two years of the time series by using the \texttt{forecast} function. The result is displayed in Fig. \ref{fig:log_airline_forecast}.
\vspace{2mm}
\begin{lstlisting}[language = Julia]
# Number of months ahead to be forecast
N = 24
# Perform forecasting
pred, dist = forecast(ss, N)
\end{lstlisting}

\subsection{Electricity consumption}

Another practical example is the study of monthly electricity consumption in a given area. In this example, we will use real data from a Brazilian distribution company. In addition, we will consider a temperature series, which is highly correlated to electricity consumption in Brazil, as an exogenous variable. A similar study using \texttt{StateSpaceModels.jl} was conducted in \cite{saavedra2018simulating}.

In this case, our objective is to simulate a large set of future scenarios that can be used as inputs in a stochastic optimization problem with the goal of obtaining the best contracting strategy for a distribution company. To this end, the function \texttt{simulate} will be used to perform Monte Carlo simulation. We will also use the square-root Kalman filter for this application.
\begin{lstlisting}[language = Julia]
# Specify the state-space model
model = structural(consumption, 12; X = temperature)
# Estimate the state-space model
ss = statespace(model; filter_type = SquareRootFilter)
# Number of months ahead to be simulated
N = 24
# Number of scenarios to be simulated
S = 1000
# Perform simulation
sim = simulate(ss, N, S)
\end{lstlisting}

\begin{figure}[h]
	\centering
	\includegraphics[width=0.7\columnwidth]{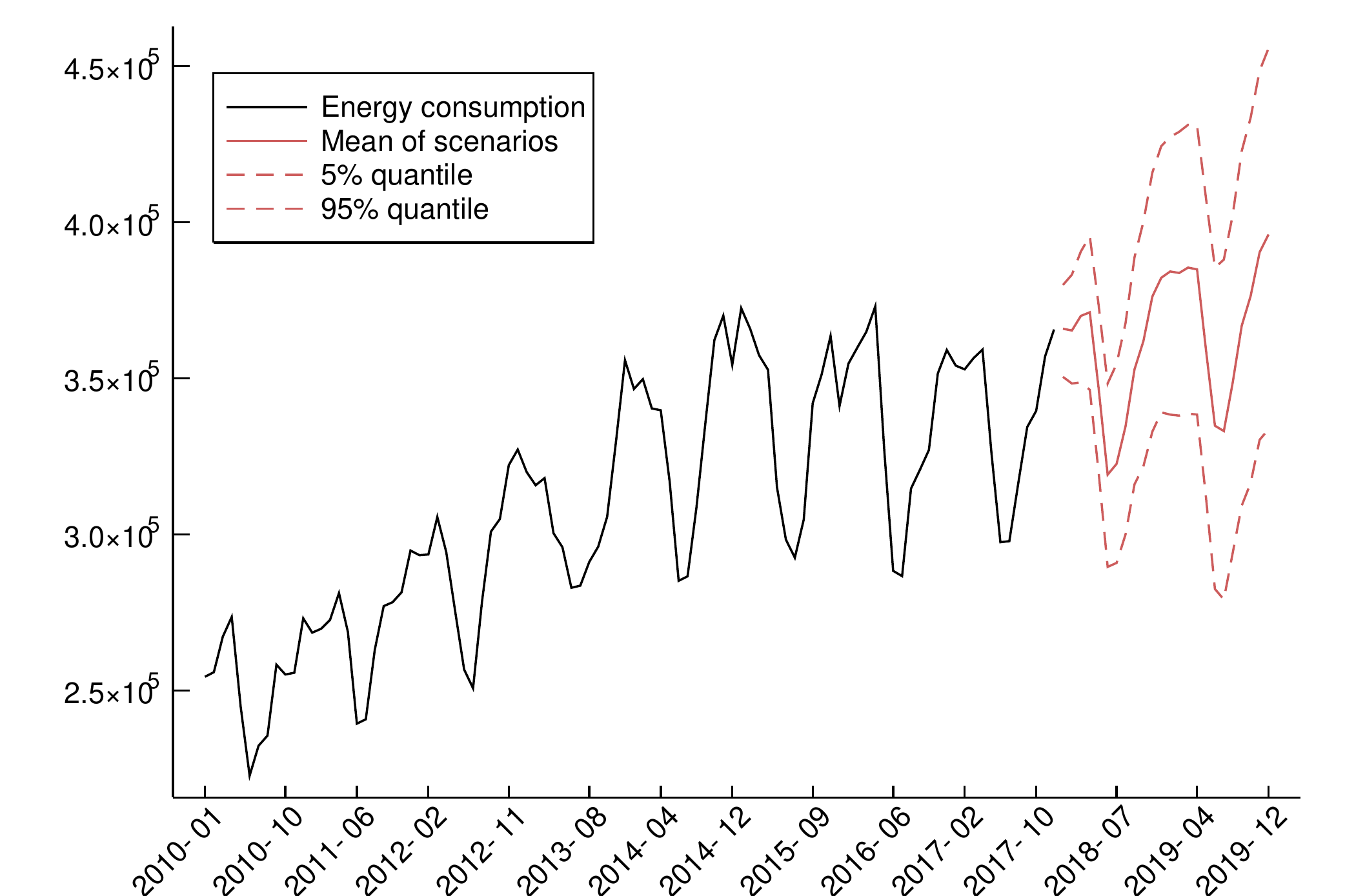}
	\caption{Monthly energy consumption historical data and simulation of future scenarios.}
	\label{fig:energy_simulation}
\end{figure}

The time series and the resulting simulation can be seen in Fig. \ref{fig:energy_simulation}. The future scenarios are graphically represented by their mean, which is identical to the forecast at each time period, as well as the 5\% and 95\% quantiles.

\subsection{Vehicle tracking}

Finally, in order to illustrate one application that does not fall into any of the predefined models, thus requiring a user-defined model, let us consider an example from control theory. More precisely, we are going to use \texttt{StateSpaceModels.jl} to track a vehicle from noisy sensor data. In this case, $y_t$ is a $2 \times 1$ observation vector representing the corrupted measurements of the vehicle's position on the two-dimensional plane in instant $t$. Since sensors collect the observations with the presence of additive Gaussian noise, we need to filter the observation in order to obtain a better estimate of the vehicle's position.

The position and speed in each dimension compose the state of the vehicle. Let us refer to $x_t^{(d)}$ as the position on the axis $d$ and to $\dot{x}^{(d)}_t$ as the speed on the axis $d$ in instant $t$. Additionally, let $\eta^{(d)}_t$ be the input drive force on the axis $d$, which acts as state noise. For a single dimension, we can describe the vehicle dynamics as 
\begin{equation}
\begin{aligned}
    & x_{t+1}^{(d)} = x_t^{(d)} + \Big( 1 - \frac{\rho \Delta_t}{2} \Big) \Delta_t \dot{x}^{(d)}_t + \frac{\Delta^2_t}{2} \eta_t^{(d)}, \\
    & \dot{x}^{(d)}_{t+1} = (1 - \rho) \dot{x}^{(d)}_{t} + \Delta_t \eta^{(d)}_t,
\end{aligned}\label{eq_control}
\end{equation}
where $\Delta_t$ is the time step and $\rho$ is a known damping effect on speed. 

We can cast the dynamical system (\ref{eq_control}) as a state-space model in the following manner:
\begin{align*} 
    y_t &= \begin{bmatrix} 1 & 0 & 0 & 0 \\ 0 & 0 & 1 & 0 \end{bmatrix} \alpha_{t+1} + \varepsilon_t, \\
    \alpha_{t+1} &= \begin{bmatrix} 1 & (1 - \tfrac{\rho \Delta_t}{2}) \Delta_t & 0 & 0 \\ 0 & (1 - \rho) & 0 & 0 \\ 0 & 0 & 1 & (1 - \tfrac{\rho \Delta_t}{2}) \\ 0 & 0 & 0 & (1 - \rho) \end{bmatrix} \alpha_{t} + \begin{bmatrix} \tfrac{\Delta^2_t}{2} & 0 \\ \Delta_t & 0 \\ 0 & \tfrac{\Delta^2_t}{2} \\ 0 & \Delta_t \end{bmatrix} \eta_{t},
\end{align*}
where $\alpha_t = (x_t^{(1)}, \dot{x}^{(1)}_{t}, x_t^{(2)}, \dot{x}^{(2)}_{t})^{\top}$ and $\eta_t = (\eta^{(1)}_t, \eta^{(2)}_t)^{\top}$.

\vspace{0.3cm}

We can formulate the vehicle tracking problem in the \texttt{StateSpaceModels.jl} framework as:

\begin{lstlisting}[language = Julia]
# State transition matrix
T = kron(Matrix{Float64}(I, p, p), [1 (1 - ρ * Δ / 2) * Δ; 0 (1 - ρ * Δ)])
# Input matrix
R = kron(Matrix{Float64}(I, p, p), [.5 * Δ^2; Δ])
# Output (measurement) matrix
Z = kron(Matrix{Float64}(I, p, p), [1 0])
# User defined model
model = StateSpaceModel(y, Z, T, R)
# Estimate vehicle speed and position
ss = statespace(model)
\end{lstlisting}

In this example, we define the noise variances $H$ and $Q$, generate the noises and simulate a random vehicle trajectory using the state-space equations:
\begin{lstlisting}[language = Julia]
# Generate random actuators
Q = .5 * Matrix{Float64}(I, q, q)
η = MvNormal(zeros(q), Q)
# Generate random measurement noise
H = 2. * Matrix{Float64}(I, p, p)
ε = MvNormal(zeros(p), H)
# Simulate vehicle trajectory
α = Matrix{Float64}(undef, n + 1, m)
y = Matrix{Float64}(undef, n, p)
for t in 1:n
    y[t, :] = Z * α[t, :] + rand(ε)
    α[t + 1, :] = T * α[t, :] + R * rand(η)  
end
\end{lstlisting}

\begin{figure}[h]
	\centering
	\includegraphics[width=0.7\columnwidth]{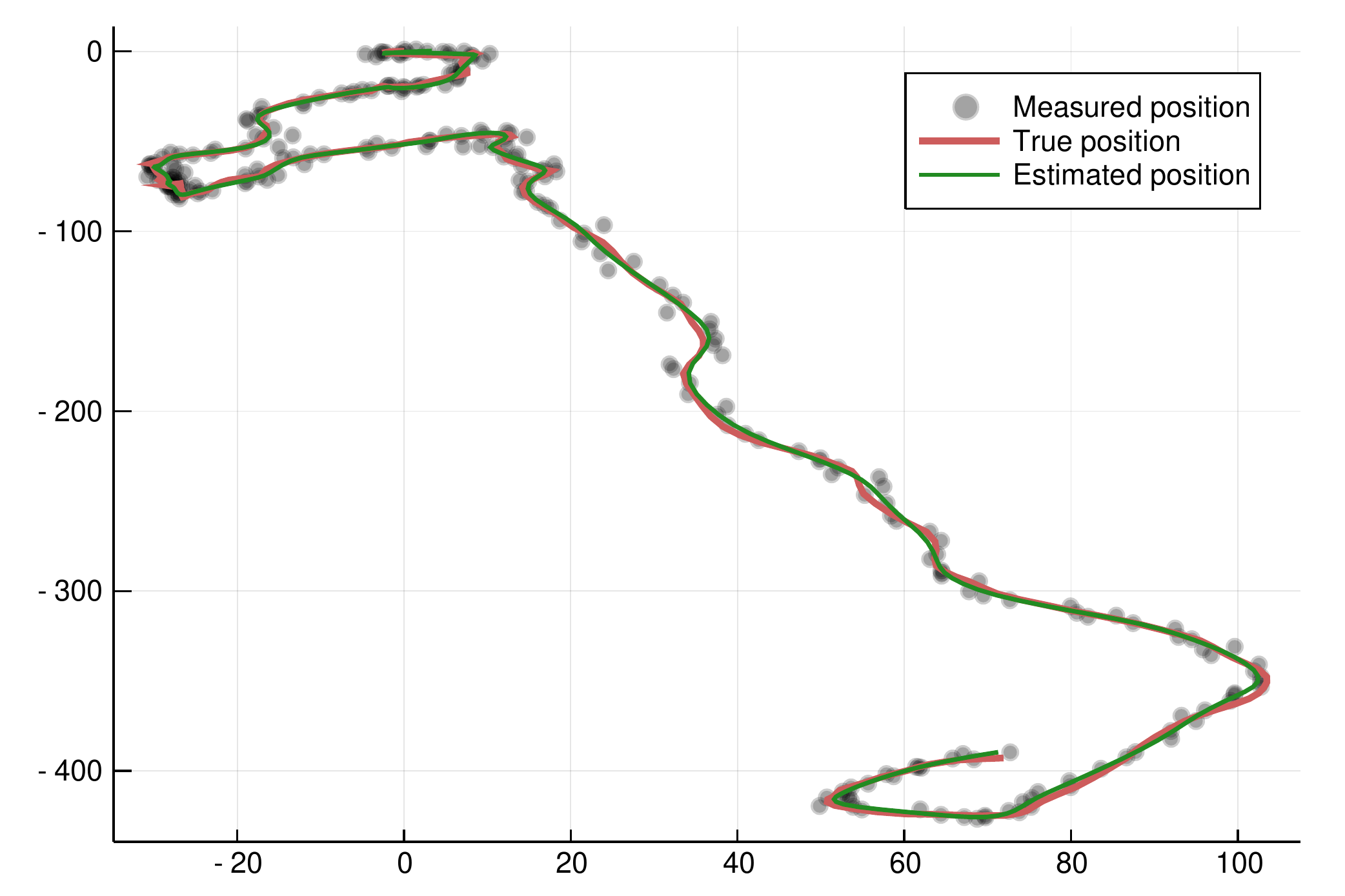}
	\caption{Vehicle tracking using a state-space model.}
	\label{fig:vehicle_tracking}
\end{figure}

An illustration of the results can be seen in Fig. \ref{fig:vehicle_tracking}. It can be seen that the measurements are reasonably noisy when compared to the true position. Furthermore, the estimated positions, represented by the smoothed state, effectively estimate the true positions with small inaccuracies. 

\section{Conclusion} \label{sec:conclusion}

\texttt{StateSpaceModels.jl} is a flexible package for time-series modeling, forecasting, and simulating that is fully implemented in Julia. The package contains an implementation of the Kalman filter and smoother as well as their square-root variants. Additionally, users have the ability to use any implemented filter or optimization method. Missing observations in the form of \texttt{NaN} values are automatically treated in the filtering and smoothing steps. 

Besides comprising several predefined classical models, it is also possible to define any linear model with the \texttt{StateSpaceModel} constructor. Forecasting and Monte Carlo simulation of future scenarios are also available. Finally, the package documentation contains a manual with straightforward examples that are simple to reproduce.

\bibliographystyle{IEEEtran}
\bibliography{ref}

\end{document}